# QoE-Based Scheduling Algorithm in WiMAX Network using Manhattan Grid Mobility Model

Tarik ANOUARI
Computer, Networks, Mobility and Modeling laboratory
Department of Mathematics and Computer
FST, Hassan 1st University, Settat, Morocco

Abdelkrim HAQIQ
e-NGN Research group, Africa and Middle East

Abstract— WiMAX (acronym for Worldwide Interoperability for Microwave Access) is a family of technical standards based on IEEE 802.16 standard that defines the high speed connection through radio waves unlike DSL (Digital Subscriber Line) or other wired technology. It can provide coverage to remote rural areas of several kilometers in radius, it's an adequate response to some rural or inaccessible areas. WiMAX can provide point-to-point (P2P) and point-to-multipoint (PMP) modes. In parallel, it was observed that, unlike the traditional assessment methods for quality, nowadays, current research focuses on the user perceived quality, the existing scheduling approaches take into account the quality of service (QoS) and many technical parameters, but does not take into account the quality of experience (QoE). In this paper, we present a scheduling algorithm to provide QoE in WiMAX network under Manhattan Mobility. A new approach is proposed, particularly for the Best Effort (BE) service class WiMAX, in this approach, if a packet loss occurs on a link connection, the system then reduces the transmission rate of this connection to obtain its minimum allowable transmission rate. The NS-2 simulation results show that the QoE provided to users is enhanced in terms of throughput, jitter, packet loss rate and delay.

## I. INTRODUCTION

The network has been objectively examined by performing a number of measures to evaluate the quality of network service. This assessment is known as the QoS of the network. The term QoS refers to the capacity of a network to transfer in good conditions a given type of traffic, in terms of availability, throughput, transmission delay, jitter, packet loss rate ... The QoS does not take into consideration the user's perception of the quality of the service provided. Another assessment which takes into account the user's perception is named QoE, it's a subjective measure that implicates human dimensions; it groups together user perception, expectations, and experience of application and network performance.

QoE has become a very active area of research. Many related works were published on analyzing and improving QoE [14] over WiMAX network. The study reported in [16] proposes an estimation method of QoE metrics based on QoS metrics in WiMAX network. The QoE was estimated by using a Multilayer Artificial Neural Network (ANN).The results show an efficient estimation of QoE metrics with respect to QoS parameters.

Other works like [8, 9 and 10] also focus on the ANN method to adjust the input network parameters to get the ideal output to obtain the users' satisfaction. Principally, the success of the ANN approach depends on the model's capacity to completely learn the nonlinear interactions between QoE and QoS.

In [3], we proposes a QoE-Based Scheduling Algorithm to control and adapt the packets transmission rate so as to reduce packet loss rate, jitter and delay, our study focuses on UGS service class. This paper takes into account the BE WiMAX service class using the Manhattan Mobility Model.

The rest of this paper is organized as follows. Section 2 gives a short description on the Manhattan Mobility Model, WiMAX technology is described in Section 3. In Section 4, a QoE overview background is presented. The proposed QoE model is described in detail in Section 5. Simulation environment and performance parameters are described in Section 6. Section 7 shows simulation results and analysis. Finally, the paper ends with conclusion and a presentation of future work directions.

## II. MOBILITY MODELS

### A. Manhattan Mobility Model

The Manhattan Mobility Model (MMM) [4] makes use of a map to confine movement on roads (Figure 1). The node is free to move on horizontal and vertical two-lane streets. Arriving to an intersection, it can turn left or right with probability equal to 0.25 or go straight with probability 0.5. Moreover, nodes move according to a temporal correlation. The node's speed is constrained by the speed of the front node in the same lane.

Figure 1 shows the movement of a node using the Manhattan mobility model.

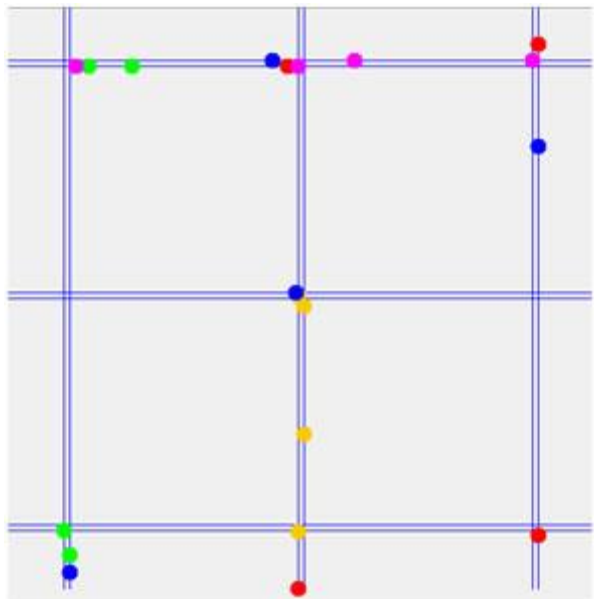

Figure 1: Nodes' movement in Manhattan Mobility Model

### III. WIMAX TECHNOLOGY

WiMAX is a set of technical standards based on the 802.16 standard [12, 13]. It uses multiple broadcasting wireless technologies mainly for PMP architecture: one or more centralized transmitters / receivers cover a area where there are multiple terminals.

WiMAX can be used in PMP mode, in which from a central base station, serving multiple client terminals is ensured and in P2P mode, in which there is a direct link between the central base station and the subscriber.

PMP mode is less expensive to implement and operates while P2P mode can provide greater bandwidth.

#### A. QoS in WiMAX Network

WiMAX standards described in detail the tools to set up a network of high-speed data but let the choice to implement their own mechanisms for QoS management. It may satisfy QoS requirements for a wide range of services and data applications especially with the high speed connection, asymmetric capabilities UL & DL and flexible mechanisms for resource allocation. Some applications like video streaming and VoIP require a short response time and cannot tolerate congestion in term of throughput, transmission delay, jitter and packet loss.

QoS is the capacity of transmission in good conditions of a number of packets in a connection between a transmitter and a receiver, and it can be presented in several terms such as availability, throughput, transmission delay, jitter, packet loss rate. The concept of QoS obviously depends on the service considered, its requirement of response time, which is its sensitivity to transmission errors... etc.

The objective of QoS is to optimize network resources and ensure good performance to applications. It can provides to users differentiated throughputs and response times according to the protocols implemented at the network layer.

Respecting QoS requirement becomes very important in IEEE802.16 systems to guarantee their performance, in particular in the presence of various kinds of connections, namely the current calls, new calls and the handoff connection.

#### B. Different Service Classes in WiMAX

Base stations and terminals use a service flow with an appropriate QoS class (in addition to other parameters such as bandwidth and delay) to ensure the appropriate QoS treatment to the application. Each connection is associated with a single data service. Each data service is associated with a set of QoS parameters that measure its behavior. To satisfy different kinds of applications, WiMAX standard has defined four service classes of quality, namely BE, Real-Time Polling Service (rtPS), Non-Real Time Polling Service (nrtPS) and unsolicited Grant Service (UGS). The ertPS service class was added specifically for the mobile version [1].

Table 1 classifies different service classes of WiMAX and gives their description and QoS parameters.

TABLE 1: SERVICE CLASSES IN WiMAX

| Service | Description | QoS parameters |
|---------|-------------|----------------|
| UGS | Real-time data streams comprising fixed size data packets at periodic intervals | Maximum Sustained Rate<br>Maximum Latency Tolerance<br>Jitter Tolerance |
| rtPS | Support real-time service flows that periodically generate variable-size data packets | Traffic priority<br>Maximum latency tolerance<br>Maximum reserved rate |
| ertPS | Real-time service flows that generate variable-sized data packets on a periodic basis. | Minimum Reserved Rate<br>Maximum Sustained Rate<br>Maximum Latency Tolerance<br>Jitter Tolerance<br>Traffic Priority |
| nrtPS | Support for non-real-time services that require variable size data grants on a regular basis | Traffic priority<br>Maximum reserved rate<br>Maximum sustained rate |
| BE | Data streams for which no data minimum service level is required. | Maximum Sustained Rate<br>Traffic Priority |

### IV. QUALITY OF EXPERIENCE

Quality of Experience (QoE) is a subjective assessment of a customer's experiences with a service, it focuses on the entire service, and it involves subjective human perception. QoE is in part related to QoS and are two complementary concepts.

#### A. Quality of Experience vs Quality of Service assessment

QoS and QoE are rather unclear terms sometimes used interchangeably. It is good to redefine the terms of QoS and QoE. QoE is related to but differs from QoS, which attempts to objectively evaluate the service provided by the vendor, with QoS measurement is most of the time not related to customer, but to hardware and / or software. QoS appeared in the 90 years to describe a set of techniques to ensure the good delivery of sensitive network traffic such as voice or applications. But with the rapid evolution of multimedia applications, the metrics of the QoS such as bandwidth, delay, jitter and packet loss fail to assess subjectivity associated with human perception and thus was born the QoE, which is a measure of personal judgment of the user according to his experience. Indeed, the notion of user experience has been introduced for the first time by Dr. Donald Norman, citing the importance of designing a user [18] centered service.

Gulliver and Ghinea [11] decompose QoE into three components: assimilation, judgment and satisfaction. The assimilation is a quality measure of the clarity of the contents by an informative point of view. The judgment of quality reflects the quality of presentation. Satisfaction indicates the degree of overall assessment of the user.

*B. QoE Measurement approaches*

Methods for assessing the quality perceived by the user are based on two main methods, namely, the subjective and the objective assessment. Objective methods are mainly based on mathematical algorithms that generate a quantitative measure of the service provided. While subjective assessments are carried out by human subjects to measure the overall perceived quality in a controlled environment. The most frequently used assessment is the MOS recommended by the International Telecommunication Union (ITU) [15] defined as a numeric value evaluation from 1 to 5 (i.e. poor to excellent).

Peter and Bjørn [7] categorized the existing approaches of measuring network service quality from a user perception into three classifications, namely: Testing User-perceived QoS (TUQ), Surveying Subjective QoE (SSQ) and Modeling Media Quality (MMQ). The first two approaches collect subjective information from users, whereas the third approach is based on objective technical assessment. Figure 2 [2] gives an overview of the classification of the existing approaches.

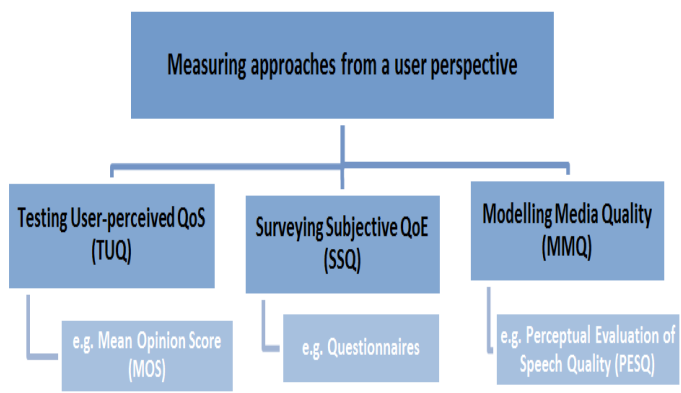

Figure 2: The approaches for measuring network service quality from a user perception

V. QOE-BASED SCHEDULING ALGORITHM MODEL

Our scheduling algorithm presented in this section can provide QoE to the WiMAX network, while the existing scheduling algorithms take into account QoS but do not provide QoE. Indeed, every user has different subjective requirement of the system.

*A. Proposed QoE Model*

We propose a QoE model in which we use two QoE requirements; an initial maximum transmission rate and a minimum subjective rate requirement for each user. Each node starts traffic with a maximum rate. When a packet loss occurs over a given connection then the system verify on each user if the transmission rate is higher than the minimum subjective rate requirement, in this case the transmission rate is reduced, otherwise it's remained at the same level. The rate returns to the original maximum value during the simulation.

Figure 3 shows the activity diagram of the proposed model

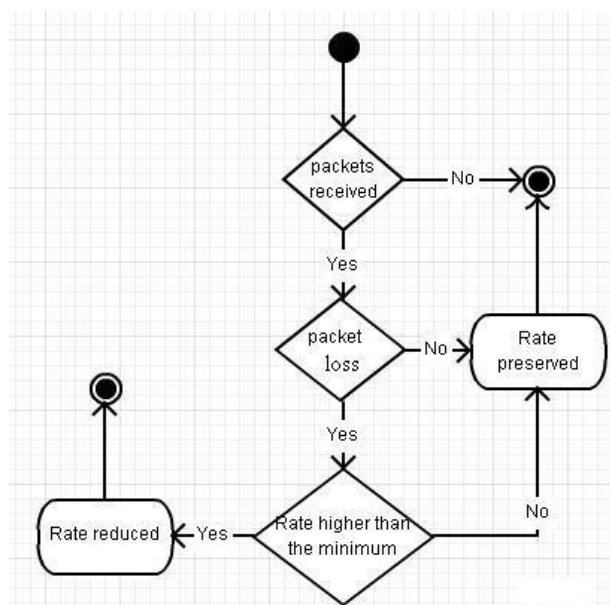

Figure 3: Activity diagram of the proposed QoE-Model

VI. SIMULATION ENVIRONNEMENT

*A. Simulation Model*

In this paper, we analyze the performances of the proposed QoE-based scheduling algorithm using the Manhattan mobility Model, our QoE-model is compared with the WiMAX module developed by NIST (National Institute for Standards and Technologies) based on the IEEE 802.16 standard (802.16-2004) and the mobility extension (80216e-2005) [17], it provides a number of features including OFDM PHY layer. Five wireless nodes (SS, subscriber stations) are created and connected to a base station (BS). A sink node is created and attached to the base station to accept incoming packets. A traffic agent is created and then attached to the source node. The Network Simulator (NS-2) [19] is used.

Finally, we set the traffic that produces each node. The first node has run with CBR (Constant Bit Rate) packet size of 200 bytes and interval of "0,0015", the second node has run with CBR packet size of 200 bytes and interval of "0,001", the third node has run with CBR packet size of 200 bytes and interval of "0,001", the fourth node has run with CBR packets size of 200 bytes and interval of "0,001" and the fifth node has run with CBR packet size of 200 bytes and interval of "0,0015". The initial transmission rate that produces each node is about "133,3 Kbps", "200 Kbps", "200 Kbps", "200 Kbps" and "133,3 Kbps" respectively. All nodes have the same priority.

Each user has a minimum requirement, so the first user requires minimal traffic rate of "120 Kbps", the second "150 Kbps", the third "150 Kbps", the fourth "150 Kbps" and the fifth "120 Kbps".

The following table summarizes the above description about the produced and required traffic rate of each user.

TABLE 2: USER'S TRAFFIC PARAMETERS

| Traffic rate / Users | Initial traffic rate (Kbps) | User minimum requirement (Kbps) |
|---|---|---|
| User 1 | 133,33 (200byte/0. 0015) | 120 |
| User 2 | 200 (200byte/0. 001) | 150 |
| User 3 | 200 (200byte/0. 001) | 150 |
| User 4 | 200 (200byte/0. 001) | 150 |
| User 5 | 133.33 (200byte/0. 0015) | 120 |

To perform this simulation, we have implemented the QoS-included WiMAX module [5] within NS-2.29 Simulator. This module is based on the NIST implementation of WiMAX [17], it consists of the addition of the QoS classes as well as the management of the QoS requirements, unicast and contention request opportunities mechanisms, and scheduling algorithms for the UGS, rtPS and BE QoS classes.

We have generated mobility scenarios for Manhattan Grid Model using the BONNMOTION [6] tool and have converted generated scripts to the supported NS2 format so that they can be integrated into TCL scripts.

We have not been able to exceed 5 mobile nodes due to the limited performance of available computers.

Mobility models were created with speed of 15 m/s and simulation time of 200secs.

We have used PERL script to extract data from trace files in term of throughput, packet loss rate, jitter and delay. The extracted analysis results are plotted in graphs using EXCEL software.

*B. Simulation Parameters*

Simulation parameters are shown in table 3:

TABLE 3: SIMULATION PARAMETERS

| Parameter | Value |
|---|---|
| Simulator | NS-2 (Version 2.29) |
| Network interface type | Phy/WirelessPhy/OFDM |
| Propagation model | Propagation/OFDM |
| MAC type | Mac/802_16/BS |
| Antenna model | Antenna/OmniAntenna |
| Service class | BE |
| packet size | 200 bytes |
| Frequency bandwidth | 5 MHz |
| Receive Power Threshold | 2,025e-12 |
| Carrier Sense Power Threshold | 0,9 * Receive Power Threshold |
| channel | 3,486e+9 |
| Mobility Model | ManhattenGrid |
| Speed | 15 m/s |
| Simulation time | 200s |

*C. Performance Parameters*

Our simulation focuses on analyzing main QoS parameters for WiMAX Network, namely average throughput, packet loss rate, average jitter and average delay.

VII. SIMULATION RESULTS AND ANALYSIS

In this section we present the results obtained through simulations, for both traffic scenarios considered, reflecting the performance of QoE-based scheduler algorithm and the NIST scheduler in term of average throughput, packet loss rate, average delay and average jitter in WiMAX network using BE service class.

Figure 4 shows the values of the average throughput of the two modules considered in our simulations. We note that the values of the average throughput using the WiMAX module are higher than those of the module using the proposed mechanism for all the flows. Indeed, the mechanism based on QoE control the transmission rate for different users to adjust with subjective minimum requirements of each user in the main objective to reduce the network overhead and thus reduce delay, jitter and packet loss rate.

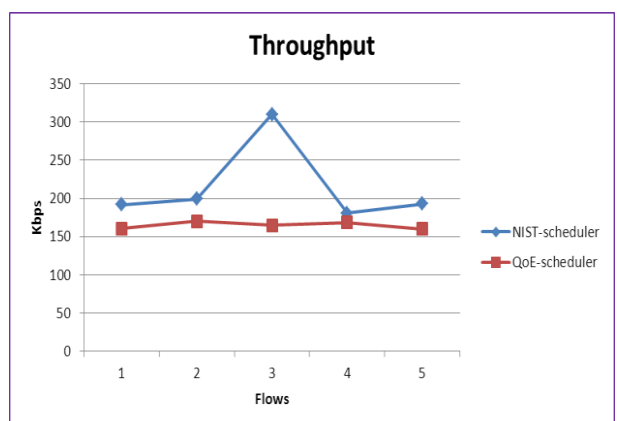

Figure 4. Average Throughput under Speed/fixed 15 m/s

Figure 5 illustrates the improvement obtained on packet loss rate by applying QoE-based scheduler algorithm for all flows. In general, the packet loss rate is reduced. In the case of flow 4, the values are similar.

136136136

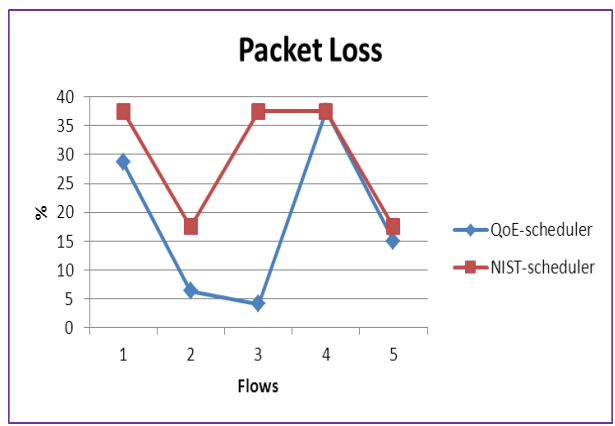

Figure 5. Packet loss rate under Speed/fixed 15 m/s

From figure 6, it can be noticed that the proposed mechanism based on the QoE is more efficient in terms of average jitter compared to the WiMAX module. Indeed, the average jitter values corresponding to the proposed mechanism are lowest compared to WiMAX module ones.

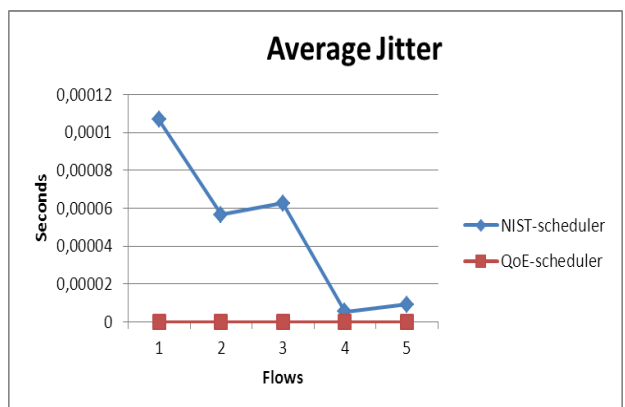

Figure 6. Average Jitter under Speed/fixed 15 m/s

As we can see on the figure 7, the average packet transmission delay is reduced using the mechanism based on QoE. In the case of flows 4 and 5, the two modules give similar average delay values.

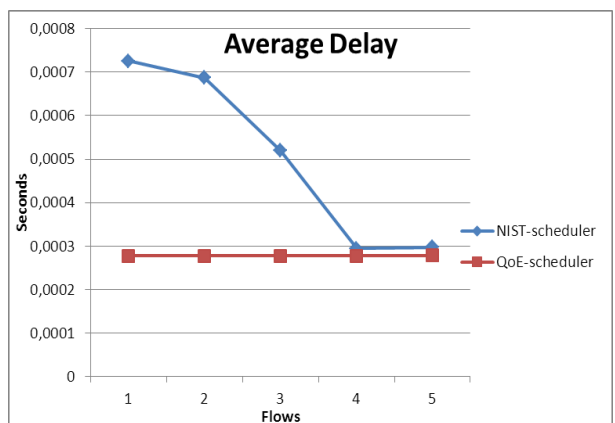

Figure 7. Average Delay under Speed/fixed 15 m/s

## VIII. CONCLUSION

In this paper, we have proposed a QoE-based mechanisms to adapt the transmission rate based on the QoE in which depending on whether there is a packet loss, the system checks for each user if the transmission rate is higher than the minimum subjective condition, if so, the transmission rate is reduced, if not it remained at the same rate.

Simulations show that the use of the proposed mechanisms improves the QoE of users in WiMAX networks. These mechanisms allow significantly reducing packet loss, jitter and delay while the transmission rate is reduced for each connection to achieve the subjective minimum transmission rate of each user and avoid congestion network.

As a future work we may extend this study by using other mobility models and taking in consideration other subjective parameters.

AUTHORS PROFILE

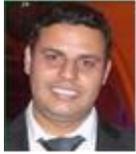

Tarik ANOUARI has a High Specialized Study Degree (DESS) option Information Systems Engineering, from the University of Cadi Ayyad, Faculty of Sciences Semlalia, Marrakesh, Morocco. Since November 2006 he has been working as Engineer Analyst Developer in the Deposit and Management fund (CDG), Rabat, Morocco. Currently, he is working toward his Ph.D. at the Faculty of Sciences and Techniques, Settat. His current research interests Simulation Network Performance, Network Protocols, Mobile Broadband Wireless and Analysis of Quality of experience in Next Generation Networks.